\documentclass[%
 reprint,
 amsmath,amssymb,
 aps,
]{revtex4-1}

\usepackage{graphicx}% Include figure files
\usepackage{dcolumn}% Align table columns on decimal point
\usepackage{bm}% bold math
\begin{document}

\preprint{APS/123-QED}

\title{Designing Magnetic Topological van der Waals Heterostructure}% Force line breaks with \\

\author{Anh Pham}
 \email{phamad@ornl.gov}
\author{Panchapakesan Ganesh}%
 \email{ganeshp@ornl.gov}
\affiliation{Center for Nanophase Materials Sciences, Oak Ridge National Laboratory, Oak Ridge, Tennessee 37831, USA}

\date{\today}% It is always \today, today,

\begin{abstract}
We demonstrate a new method of designing 2D functional magnetic topological heterostructure (HS) by exploiting the vdw heterostructure (vdw-HS) through combining 2D magnet CrI$_3$ and 2D materials (Ge/Sb) to realize new 2D topological system with nonzero Chern number (C=1) and chiral edge state. The nontrivial topology originates primarily from the CrI$_3$ layer while the non-magnetic element induces the charge transfer process and proximity enhanced spin-orbit coupling. Due to these unique properties, our topological magnetic vdw-HS overcomes the weak magnetization via proximity effect in previous designs since the magnetization and topology coexist in the same magnetic layer. Specifically, our systems of bilayer CrI$_3$/Sb and trilayer CrI$_3$/Sb/CrI$_3$ exhibit different topological ground state ranging from antiferromagnetic topological crystalline insulator (C$_M$= 2) to a QAHE. These nontrivial topological transition is shown to be switchable in a trilayer configuration due to the magnetic switching from antiferromagnetism to ferromangetism in the presence an external perpendicular electric field with value as small as 0.05 eV/\AA. Thus our study proposes a realistic system to design switchable magnetic topological device with electric field.
\end{abstract}

\maketitle

%\tableofcontents

\section{\label{sec:level1}Introduction}
Two dimensional magnetic topological materials represent a novel platform which exhibit the quantum anomalous Hall effect (QAHE) [1,2]. To realize the quantum anomalous Hall (QAH) state, the material system needs to break time reversal symmetry while maintaining a nontrivial 2D bulk gap protected by a nonzero Chern number. As a result, different pathways with an emphasis on inducing magnetization on a nontrivial topological material have been proposed. Specifically adatoms [3-6], impurity doping [7-9], interfacing between magnetic substrates and topological materials [10-16] have been proposed as possible strategy to realize a room-temperature QAHE. However, these systems require the incorporation of external impurity or the interfacing with a non 2D material which can be experimentally challenging. Alternatively, the material discovery approach has also suggested various 2D materials with large spin orbit interaction [17-19] as a mean to achieve a high-temperature QAHE, but this method also introduces new challenges in terms of material fabrication, processing, and characterization. \\
\indent Two dimensional (2D) vertical van der waals heterostructures have emerged as a new functional material class with flexible functional properties since the heterostructure can exhibit new quantum properties which are absent in the individual layers [20]. These structures require relatively low cost fabrication method which allows the combination of different materials with tailored electronic property in various vertical architecture. Consequently, 2D-HS represents an ideal platform to realize the QAHE by combining 2D materials with large spin-orbit coupling (SOC) and magnetic interaction. Among many 2D materials, CrI$_3$ is a new class of magnetic semiconductor which has demonstrated tunable exchange coupling with high Curie temperature of 45 K in the ferromagnetic phase [21-26]. In addition, different heterostructure configuration of CrI$_3$ and other material like graphene [27, 28] or Bi$_2$Se$_3$ [29] have been proposed to realize the QAHE. \\
\indent A principle design of many magnetic topological vdw-HS is the proximity effect in which the magnetization is induced on a non-magnetic but topologically non-trivial material [27-29]. However, the drawback of this approach is that the magnetic moments on the topological insulator (TI) tend to be small since most topological materials are primarily composed of materials with delocalized $\sl p$-orbital, while most magnetic materials predominantly originate from elements with half-filled $\sl d$ bands. CrI$_3$ monolayer is a unique magnetic material which has nontrivial topology with high Chern numbers inside the conduction and valence band. As a result, it has been suggested that the nontrivial topology and strong magnetization can be exploited by extreme doping up to one electron(hole)/unit cell to move the Fermi level inside the conduction or valence bands [30]. This high level of doping is unfeasible for experimental synthesis, but the Fermi level can be tuned significantly in a heterostructure configuration by creating large band offsets. \\
\indent Motivated by this idea, we first search for combination of 2D monolayer materials and CrI$_3$ as shown in table I to design a heterostructure with the lowest strain. Since Ge and Sb both have larger SOC than Si and minimal lattice mismatch with CrI$_3$, we focus our study on the different vertical vdw-HS of these materials. In addition, germanene is (Ge) a topological insulator [31], while antimonene is a normal insulator, which can be turned into a topological insulator through a large tensile strain [32]. In these hetereostructures, the individual layers can induce different charge transfer process thus resulting different bands contributing the Fermi level. In addition to the charge transfer process, other mechanisms like proximity enhanced SOC and spin transfer between the magnetic and non-magnetic layer, can further complicate the topological properties in the heterostructure. In our study,  by characterizing the electronic, magnetic and topological properties of the different layers in the heterostructure, we demonstrate that the bilayer Sb/CrI$_3$ and trilayer CrI$_3$/Sb/CrI$_3$ configurations posses nontrivial topological properties with Chern number C= $\pm1$. These topological property originates primarily from the CrI$_3$ layer which is electron doped from the elemental Sb monolayer thus resulting in QAHE with edge states close to the Fermi level. While system with weak charge transfer like Ge/CrI$_3$ or CrI$_3$Ge/CrI$_3$ remain trivial metal. The Sb also induces large SOC on the CrI$_3$ layer through proximity effect thus opening up a possibility for high temperature QAHE. In addition, the trilayer system exhibits a tunable topological transition from an antiferromagnetic topological crystalline insulator (TCI) when the CrI$_3$ layers couple antiferromagnetically to a QAHE when CrI$_3$ layers are ferromagnetically aligned. We show that this magnetic transition can be induced by a small perpendicular electric field of 0.05 eV/\AA, consitant with recent experiments on bilayer CrI$_3$. Such a value has been recently realized in a monolayer and bilayer Na$_{3}$Bi system [32] to induce the topological phase transition. Consequently, our results demonstrate a realistic platform of controlling a magnetic topological system which is suitable for experimental realization.
\begin{table}[h!]
\centering
\caption{Lattice constants and strain of different 2D monolayer elements and CrI$_3$}
\label{table:I}
\begin{tabular}{c c c c} 
 \hline\hline
 2D materials & Lattice(\AA) & vdw-HS & Lattice mismatch(\%) \\
 \hline
 Si & 3.87 & CrI$_3$/Si/CrI$_3$ & -3.13 \\ 
 Ge & 4.04 & CrI$_3$/Ge/CrI$_3$ & 1.3 \\
 Sb & 4.12 & CrI$_3$/Sb/CrI$_3$ & 3.11 \\
 CrI$_3$ & 6.91 & N/A & N/A \\ [1ex] 
 \hline
\end{tabular}
\end{table}
\section{\label{sec:level2}Computational methods}
\begin{figure*}[t]
  \centering
  \includegraphics[width=0.5\textwidth]{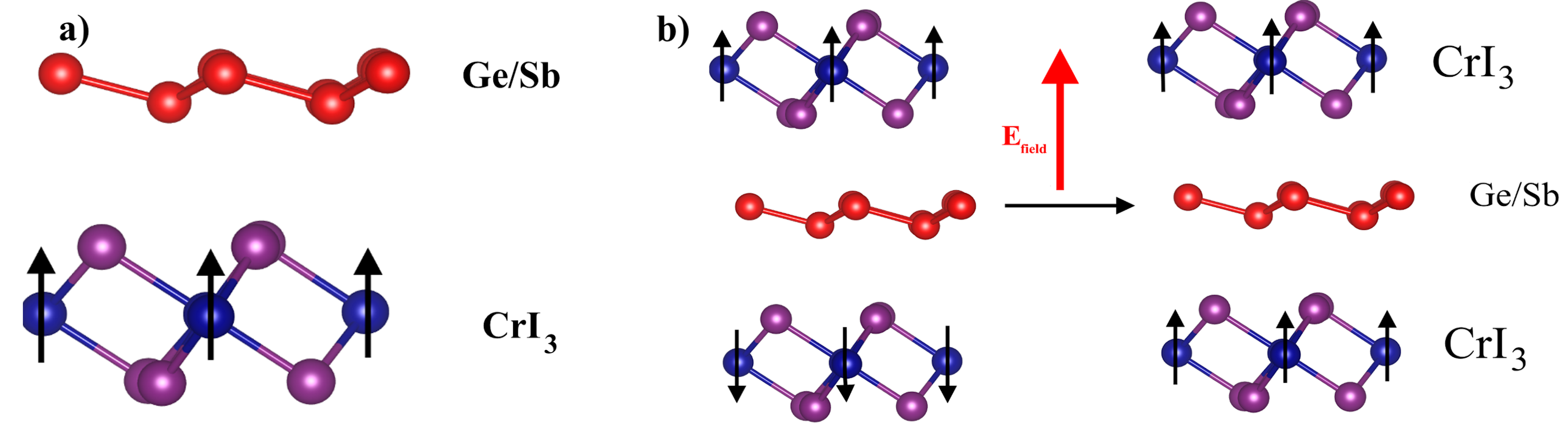}
  \caption{Schematic illustration of different CrI$_3$ based vdw-heterostructuctues with Ge/Sb. a) Bilayer. b) Trilayer}
\end{figure*}
The first principle calculations were done using the VASP package with the projected augmented wave method (PAW) [33] and the Perdew–Burke-Ernzerhof functional [35]. An energy cutoff of 600 eV was used for the structural relaxation and the electronic calculations. The DFT+U [36] with U=3.0 eV and J=0.9 eV was used in Cr's $\sl d$ to account for the correlation effect. For the individual monolayers, a dense k-point mesh of 21 $\times$ 21 $\times$ 1 was used to determine the equilibrium lattice constants, while a k-mesh of 9 $\times$ 9 $\times$ 1 was used for the structural relaxation of the heterostructure.  To avoid the interaction between the image layers in the monolayers and heterostructure configurations, vacuum was included on the top and bottom of the configuration with an overall thickness of 20 \AA. In the vdw-HS structures, the in-plane lattice constants were set at the values of the monolayers of Ge/Sb so that the lattice mismatch between CrI$_3$ and the monolayers do not affect their topology. To reduce the lattice mismatch, a configuration $\sqrt{2}$ $\times$ $\sqrt{2}$ was used for the mono elemental layers. The geometrical structures of the vdw-HS are summarized in Table I. The magnetic ground state in the trilayer structure was studied by setting the interlayer coupling between the CrI$_3$ layers to be ferromagnetic (FM) and antiferromagnetic (AFM). The spin-orbit coupling was included in a self-consistent approach to determine the nontrivial topological property. To calculate the edge state in the magnetic topological phase, a Wannier tight binding method [37] was used to obtain the Green's function [38] spectrum as implemented in the WannierTools code. The Wannier basis was calculated from the Wannier90 codes [39] with the orbital characteristic of Cr's $\sl d$ orbitals, I's $\sl p$ orbitals, Ge/Sb and Te's $\sl p$ orbitals.
\section{\label{sec:level3}Results and Discussions}
\begin{figure*}[t]
  \centering
  \includegraphics[width=1.0\textwidth]{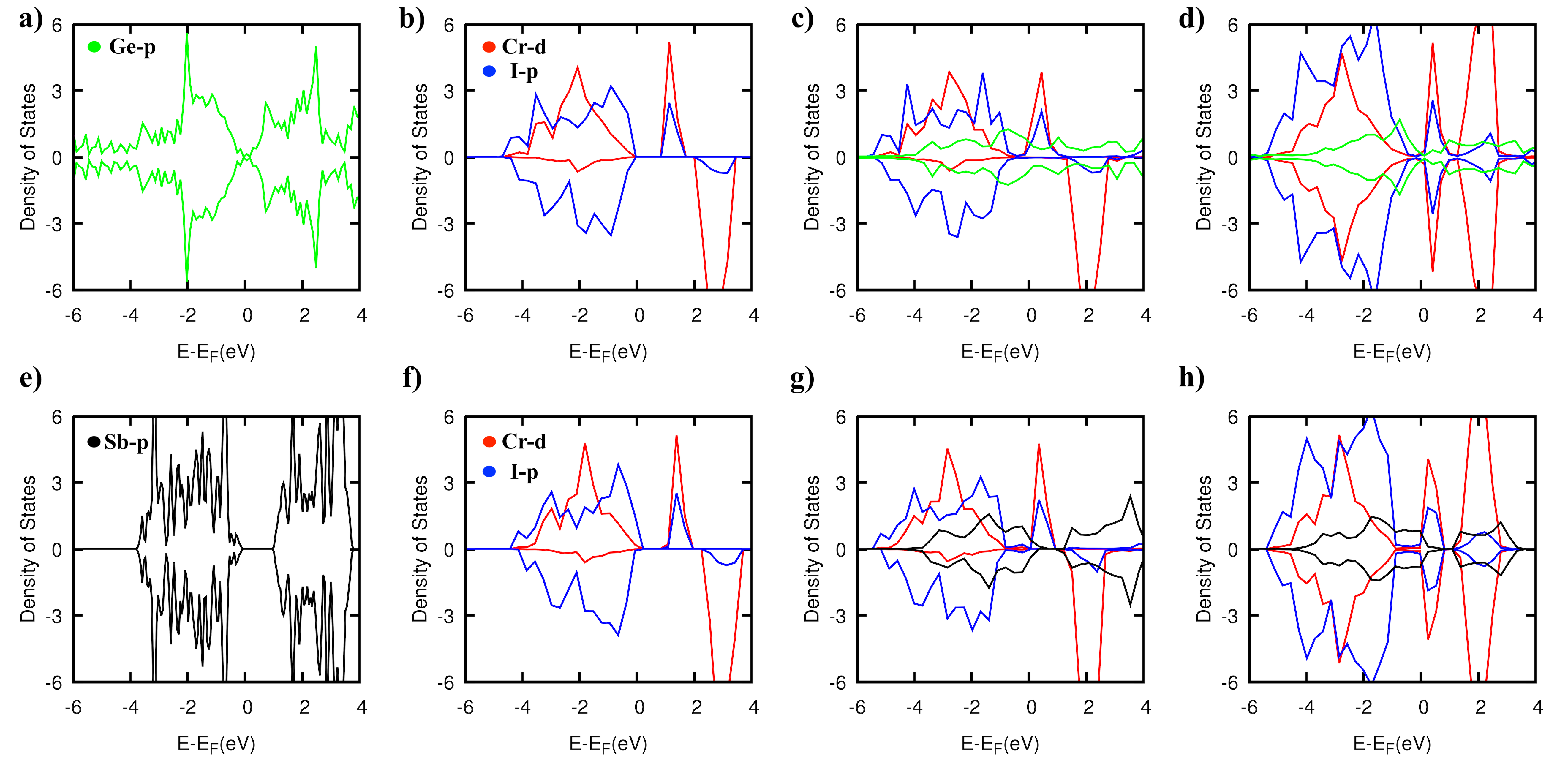}
  \caption{a), (e) Density of states of monolayer Ge and Sb. (b), (f) Partial density of states of Cr-d and I-p in monolayer ferromagnetic CrI$_3$ calculated with Ge and Sb in-plane lattice respectively. (c), (g) Partial density of states of Cr-d,I-p and Ge-p (Sb-p) in ferromagnetic CrI$_3$/Ge(Sb) heterostructure respectively. (d),(h) Partial density of states of Cr-d, I-p and Ge-p in antiferromagnetic CrI$_3$/Ge(Sb)/CrI$_3$ trilayer respectively.} 
\end{figure*}
\begin{figure*}[t]
  \centering
  \includegraphics[width=1.0\textwidth]{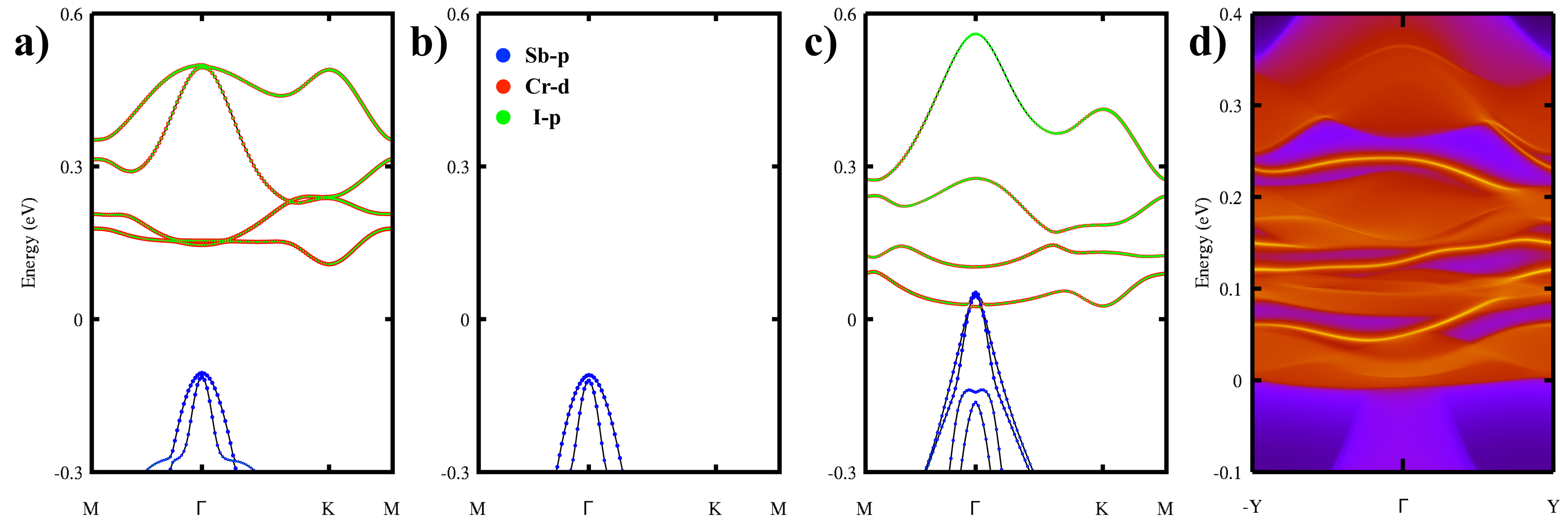}
  \caption{Electronic and topological property of bilayer Sb-CrI3 heterostructure. a) Band structure of spin-up states. b) Band structure of spin down states. c) Band structure with SOC. d) Edge state along (110) direction.} 
\end{figure*}
\begin{figure*}[t]
  \centering
  \includegraphics[width=1.0\textwidth]{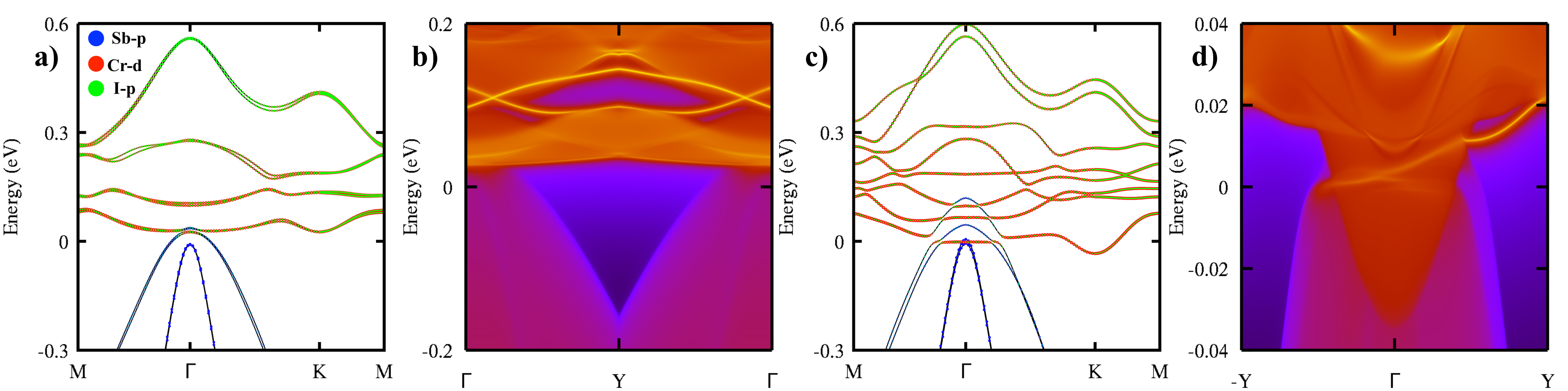}
  \caption{Electronic and topological property of trilayer CrI$_3$/Sb/Cr$_3$ heterostructure without and with perpendicular electric filed. a) Band structure wit SOC without electric field in the AFM coupling. b) Topological edge state along the (110) edge state in the AFM coupling.  c) Band structure with SOC in the FM coupling with E = 0.05 eV/\AA. d) Topological edge state along the (110) edge state in the FM coupling with E = 0.05 eV/\AA.} 
\end{figure*}
\indent The heterostructuring process can significantly modify the electronic properties of both the CrI$_3$ as well as the Ge/Sb layers. As shown in Fig. 2b and 2f, the ferromagnetic monolayer CrI$_3$ maintains the half-semiconducting groundstate even in the presence of small tensile strain in the Sb. Consequently, a band gap is clearly visible for both of the spin-channel of CrI$_3$. Electronically, similar to previous theoretical study [40-42] our monolayer CrI$_3$'s density of state is composed of a valence band dominated by the I-$\sl p$ orbitals and the Cr's fully filled t$_{2g}$ band slightly below the I-$\sl p$ orbitals, while the conduction band is composed primarily of the empty Cr's $\sl e_g$ orbital. For the monolayer Ge and Sb, the density of state also reveals a Dirac dispersion in Ge [Fig. 2a] and a large semiconducting gap in Sb [Fig. 2e] consistent with previous ab-inito studies [30,31]. When we create a bilayer structure between CrI3$_3$ and elemental Ge(Sb), this results in a significant modification of the electronic-structure. For one, the bands undergo a rigid shift so as to pin the Fermi-level to the bottom of the conduction band as illustrated in Figs. 2c and 2g. Secondly, the gap in the spin-up channel is closed, making CrI$_3$ half-metallic. This suggests that CrI$_3$ is electron-doped. Complimentarily, the Fermi-levle is pinned to the top of the valence bande-edge in Sb and Ge.  This is reminiscent of hole-doping, and results in an increase in the density-of-states at the Fermi-level for both these systems [Figs. 2c and 2g].  This leads to a net charge-transfer from the Ge(Sb) layer to CrI$_3$, as seen in Figs. S2.  Interestingly, CrI$_3$ also induces spin-polarization to the Ge(Sb) layer -- a proximity induced magnetization-effect. Consequently, the closing of the band-gap due to this charge-transfer effect and the mixing of Cr's partially occupied conduction bands with the valence bands of Ge (Sb) gives rise to a situation where new topologies can emerge. But given that we see proximity induced magnetization in Ge(Sb) and expect proximity-induced SOC in CrI$_3$, it is not clear which states give rise to the non-trivial topology. \\
\indent The ground-state magnetic-arrangement of a CrI$_3$ bilayer is antiferromagnetic. Our magnetic calculations show that even with a layer of Ge(Sb) in our trilayer configuration, this coupling remains antiferromagnetic. Similar to the result in the bilayer structure, CrI$_3$ is electron-doped while the Ge(Sb) is hole-doped, indicating a charge-transfer between them. However, there is a key difference -- while time-reversal-symmetry (TRS) is broken in the bilayer, it is preserved in the trilayer configuration.  This is enforced by a different degree of hybdridization between the monolayer and the CrI$_3$ layer in the trilayer structure. This should result in different topologies in the bilayer and the trilayer configurations.  \\
\indent In a bilayer structure, due to the hybridization of the Ge(Sb) states, opening of the gap and time-reversal-symmetry breaking due to the ferromagnetic CrI$_3$ layer, the system could possess new interesting topologies. For example, a large hybridization gap exists between the Cr's $\sl d$ and Ge's $\sl p$ bands in the spin-up channel [Figs. S1], while the spin-down channel is dominated by Ge's $\sl p$ bands, which further exhibits a massive Dirac cone because inversion symmetry is broken in the bilayer. Spin-orbit has negligible effect on the band structure [Figs. S1]. As a result, the Chern number calculation reveals a C=0 which indicates a trivial metallic state in the Ge/CrI$_3$ bilayer. Our result is consistent with recent study of the germanene on CrI$_3$ substrate [43]. \\
\indent Different from the Cr/CrI$_3$ the Sb/CrI$_3$ bilayer structure, there is a massive SOC induced splitting for the Cr-derived conduction bands [Fig. 3(c)], indicating that it is the Sb-layer that is giving rise to a large proximity induced SOC in CrI$_3$. In conjunction with ferromagnetism, this gives rise to a quantum anomalous hall effect with C = 1 and an edge states along the (110) direction as shown in Fig. 3(d). This nontrivial topology originates primarily from the large SOC induced parity exchange between the Cr's $\sl d$ and I's $\sl p$ orbitals. Given this proximity induced SOC in CrI$_3$, and the resultant large splitting of the CrI$_3$ bands, it is pertinent to ask what topological transitions can be realized by further changing the chemical-potential to occupy more of the conduction-band. For example, there is a much larger Chern gap ~25.4 meV above the Fermi level in the energy region 0.14-0.2 eV around the K point. The topological calculation incuding states up to 0.1 eV, corresponding to a Fermi-level tuned up to this value, shows a C value of 1, indicative of a quantum anamolous hall effect (QAHE) a single edge state as shown in Fig. 3d. As a result, we demonstrate that proximity induced SOC via heterostructuring can enable us to tune the system to posses a multitude of non-trivial topologies, that could even survive high-temperatures. \\
\indent To further understand the different contribution of the different layers on the nontrivial/trivial topology in the bilayer configuration, we deconstruct the total Hamiltonian in the Wannier basis as following H$^{mn}_{total}$ = H$^{mn}_{Ge/Sb}$ + H$^{mn}_{CrI_3}$ + H$^{mn}_{Ge/Sb-CrI_3}$. The individual Hamiltonian is obtained by setting the hopping components of the other orbitals equal to zero. For instance to obtain H$^{mn}_{Ge/Sb}$ only the hoppings between the Ge/Sb's $\sl p$ orbitals are retained in total Hamiltonian while the hopping containing the orbitals on CrI$_3$ is set to zero. For the Ge-based bilayer heterostructure, the individual Ge exhibits a QAHE with C=1 as shown in Fig. S3, while CrI$_3$ is topological trivial when examining the bands up to 0.2 eV. Interestingly, when the hamiltonian contains both of the Ge's and CrI$_3$ hopping but no interlayer hopping, the system becomes topological nontrivial with C=1. When the interlayer hopping is included, a small gap is opened at the $\Gamma$ point in the band structure, but this results in overall in a topological trivial metal state. On the other hand, in the Sb-based bilayer, the Sb-only layer is a normal metal, while the CrI$_3$ only band is topological nontrivial. When the hamiltonian contains both of contribution from Sb and CrI$_3$ but without and with the interlayer hopping, the topology remains nontrivial. As a result, in the Ge-based bilayer, the interlayer interaction between Ge and CrI$_3$ destroys the nontrivial topology, while in the Sb-based bilayer the nontrivial topology is originated primarily from the CrI$_3$ layer and the Sb layer plays the role of charge transfer and the proximity induced spin-orbit coupling.\\
\indent  In the case of the trilayer heterostructure, even though the magnetic ground-state between the CrI$_3$ is antiferromagnetic, this magnetic interaction can be tuned by applying a perpendicular electric field as illustrated schematically in Fig. 1b. In the CrI$_3$ bilayers, the magnetic coupling have been shown to strongly dependent on the vdw bonding between the different layers [23, 24]. Specifically, the interlayer ferromagnetic coupling can be favourable if there is reduced distance between the interlayer Cr ions resulting in interlayer vdw bonding. As revealed in table S1, the applied external perpendicular electric field results in significant structural changes in the interlayer distance between the CrI$_3$ layers and the buckling distance in the monolayers. Specifically, the buckling parameter of Ge increases with applied E-field, while the two layers of CrI$_3$ move closer together in the case of CrI$_3$/Sb/CrI$_3$. In addition, a further examination of the charge density in the trilayer [Fig. S3] reveal significant changes in the  bonding characteristic between the monolayer and the CrI$_3$ layers in the present of the electric field. In the case of CrI$_3$/Ge/Cr$_3$, the electric field enhances the buckling of the Ge layer results in bonding between the CrI$_3$ layers and the Ge monolayer due to the charge overlapping between the Ge and I atoms [Fig. S3b]. This bonding facilitates the magnetic coupling between the interlayer Cr ions thus resulting in a ferromagnetic exchange interaction. On the other hand, since the Sb layer is highly buckled there is already a weak bonding between the I atom and the Sb atom without the external field [Fig. S3c]. As the electric field is turned on, the two CrI$_3$ layers move closer towards the Sb layer thus further enhancing the bonding between the CrI$_3$ layers and the Sb layer [Fig. S3d]. This enhanced bonding results in a switch in the magnetic ground state. Consequently, these results suggest that the external electric field act as enhancement on the bonding between the vdw layers CrI$_3$ and the Ge/Sb layer which mediates the ferromagnetic interaction between the Cr ions in the different Cr$_3$ layers.  This also leads to a significantly lower critical field for switching the magnetization in the presence of antimonene. In general, this configuration shares commonality with the bilayer configureation -- both have broken inversion and time-reversal symmetry and show similar degrees of charge-transfer.  \\
\indent In the CrI$_3$/Ge/CrI$_3$, the SOC has negligible effect on the band structure [Fig. S3] similar to the bilayer configuration, while in CrI$_3$/Sb/CrI$_3$ heterostructure the nontrivial topology exists in both of AFM and FM coupling. Without the external field, the SOC opens a up a nontrivial topological gap around K in energy region from 0.1-0.25 eV, which mainly originated from the Cr's d and I'p state [Figs. 4a]. Further investigation of the edge state along the [110] direction reveals two Dirac cones which implies a nontrivial mirror Chern number (MCN) [Fig. 4b]. Further analysis using the Wannier charge center confirm a nonzero MCN with C = 2 [Fig. S7]. With a small field of 0.05 eV/\AA, the magnetic interaction is tuned to a FM state which also facilitates a transtion to a Chern semimetallic phase. In the FM configuration, the CrI$_3$/Sb/CrI$_3$ is a semiconducting ground state [Figs. S6d and 6e] similar to the bilayer system. When the SOC is effective, a drumhead like band structure [Fig. 4c] is observed and the system exhibits a QAHE with C = 1 with a single edge state as shown in Fig. 4d. As a result, CrI$_3$/Sb/CrI$_3$ HS represents a novel vdw system in which multiple nontrivial topological state is switchable through a small electric field.\\
\section{\label{sec:level4}Conclusions}
In conclusion, we have demonstrated that the appropriate combination of 2D material with 2D magnet like CrI$_3$ in a heterostructure architecture can give rise to non-trivial magnetic topological state in different layer configurations. Different from previous study of magnetic topological heterostructure in which the topology is induced to the non-magnetic but topological nontrivial material, our hetereostructure system exploit the unique topology in CrI$_3$ by designing system in which the strong bonding between the elemental element like Sb and CrI$_3$ occurs to induce a semimetallic transition in Cr$I_3$ through charge transfer process. In addition, Sb also induces a large spin-orbit coupling on the magnetic layer due to the large intrinsic SOC and the broken inversion symmetry thus resulting in different nontrivial magnetic topologies. Specifically, our results show that the Sb-based trilayer systems represent a novel tunable topological ground state which allows a switching between an AFM topological crystalline insulator to a QAHE using a very small external electric field. Furthermore, the Sb-based bilayer also exhibits a large bulk gap of 28.6 meV with nonzero Chern number which can exhibit a high-temperature QAHE. As a result, this system represents a viable strategy for further experimental exploration of topological device based on vdw-HS through the manipulation of the spin alignment within the ferromagnetic layers. 
\section*{Acknowledgements}
A.P. and P.G. were financially supported by the Oak Ridge National Laboratory's, Laboratory Directed Research and Development project. Part of the research used resources of the National Energy Research Scientific Computing Center (NERSC), a U.S. Department of Energy Office of Science User Facility operated under Contract No. DE-AC02-05CH11231. Part of this research used resources of the Oak Ridge Leadership Computing Facility, which is a DOE Office of Science User Facility supported under Contract DE-AC05-00OR22725. 
\section*{References}
\noindent{[1] C.-X. Liu, S.-C. Zhang, and X.-L. Qi, Annu. Rev. Condens. Matter Phys. 7, 301 (2016).} \\
\noindent{[2] K. He, Y. Wang, and Q.-K. Xue, Annu. Rev. Condens. Matter Phys. 9, 329 (2018).} \\
\noindent{[3] Z. Qiao, S. A. Yang, W. Feng, W.-K. Tse, J. Ding, Y. Yao, J. Wang, and Q. Niu, Phys. Rev. B 82, 161414 (2010).} \\
\noindent{[4] J. Zhang, B. Zhao, and Z. Yang, Phys. Rev. B 88, 165422 (2013).} \\
\noindent{[5] X.-L. Zhang, L.-F. Liu, and W.-M. Liu, Sci. Rep. 3, 2908 (2013).} \\
\noindent{[6] J. Hu, Z. Zhu, and R. Wu, Nano Lett. 15, 2074 (2015).} \\
\noindent{[7] C.-X. Liu, X.-L. Qi, X. Dai, Z. Fang, and S.-C. Zhang, Phys. Rev. Lett. 101, 146802 (2008).} \\
\noindent{[8] C. W. Niu, Y. Dai, L. Yu, M. Guo, Y. D. Ma, and B. B. Huang, Appl. Phys. Lett. 99,142502 (2011).}
\noindent{[9] Q. Z. Wang, X. Liu, H. J. Zhang, N. Samarth, S. C. Zhang, and C. X. Liu, Phys. Rev. Lett. 113,147201 (2014).} \\
\noindent{[10] K. F. Garrity and D. Vanderbilt, Phys. Rev. Lett. 110, 116802 (2013).} \\
\noindent{[11] Z. H. Qiao, W. Ren, H. Chen, L. Bellaiche, Z. Y. Zhang, A. H. MacDonald, and Q. Niu, Phys. Rev. Lett. 112, 116404 (2014).} \\
\noindent{[12] H. Zhang, T. Zhou, J. Zhang, B. Zhao, Y. Yao, and Z. Yang, Phys. Rev. B 94, 235409 (2017).} \\
\noindent{[13] J. Kim, K.-W. Kim, H. Wang, J. Sinova, and R. Wu, Phys. Rev. Lett. 119, 027201 (2018).} \\
\noindent{[14] F. Katmis, V. Lauter, F. S. Nogueira, B. A. Assaf, M. E. Jamer, P. Wei, B. Satpati, J. W. Freeland, I. Eremin, D. Heiman, P. J.-Herrero , and J. S. Moodera, Nature 533, 513 (2016).} \\
\noindent{[15] X. Che, K. Murata, L. Pan, Q. L. He, G. Yu, Q. Shao, G. Yin, P. Deng, Y. Fan, B. Ma, X. Liang, B. Zhang, X. Han, L. Bi, Q.-H. Yang, H. Zhang, and K. L. Wang, ACS Nano 12, 5042 (2018).} \\ 
\noindent{[16] Z. Zanolli, C. Niu, G. Bihlmayer, Y. Mokrousov, P. Mavropoulos, M. J. Verstraete, and S. Blügel, Phys. Rev. B 98, 155404 (2018).} \\
\noindent{[17] Z. F. Wang, Z. Liu, and F. Liu, Phys. Rev. Lett. 110, 196801 (2013).} \\
\noindent{[18] L. Dong, Y. Kim, D. Er, A. M. Rappe, and V. B. Shenoy, Phys. Rev. Lett. 116, 096601 (2016). } \\
\noindent{[19] J. He, X. Li,  P. Lyua, and  P. Nachtigall, Nanoscale 9, 2246 (2017).} \\
\noindent{[20] A. K. Geim, and I. V. Grigorieva, Nature 499, 419 (2013).} \\
\noindent{[21] B. Huang, G. Clark, E. Navarro-Moratalla, D. R. Klein, R. Cheng, K. L. Seyler, D. Zhong, E, Schmidgall, M. A. McGuire, D. H. Cobden, W. Yao, D. Xiao, P. Jarillo-Herrero, and X. Xu, Nature 546, 270 (2017).}\\ 
\noindent{[22] M. A. McGuire, H. Dixit, V. R. Cooper, B. C. Sales, Chem. Mater. 27, 612 (2015).}\\
\noindent{[23] N. Sivadas, S. Okamoto, X. Xu, C. J. Fennie, and D. Xiao, Nano Lett. 18, 7658 (2018).} \\
\noindent{[24] P. Jiang, C. Wang, D. Chen, Z. Zhong, Z. Yuan, Z.-Y.  Lu, and W. Ji, Phys. Rev. B 99, 144401 (2019).} \\
\noindent{[25] B. Huang, G. Clark, D. R. Klein, D. MacNeill, E. Navarro-Moratalla, K. L. Seyler, N. Wilson, M. A. McGuire, D. H. Cobden, D. Xiao, W. Yao, P. Jarillo-Herrero, and X. Xu, Nat. Nanotechnol. 13, 544 (2018).} \\
\noindent{[26] S. Jiang, J. Shan, K. F. Mak, Nat. Mater. 17, 406 (2018).} \\
\noindent{[27] J. Zhang, B. Zhao, T. Zhou, Y. Xue, C. Ma, and Z. Yang, Phys. Rev. B 97, 085401 (2018).} \\
\noindent{[28] C. Cardoso, D. Soriano, N. A. García-Martínez, and J. Fernández-Rossier, Phys. Rev. Lett. 121, 067701 (2018).} \\
\noindent{[29] Y. Hou, J. Kim, and R. Wu, Sci. Adv. 5, eaaw1874 (2019).}\\
\noindent{[30] S. Baidya, J. Yu, and C. H. Kim, Phys. Rev. B 98, 155148 (2018).}\\
\noindent{[31] C.-C. Liu, W. Feng, and Y. Yao, Phys. Rev. Lett. 107, 076802 (2011).} \\
\noindent{[32] M. W. Zhao, X. Zhang, L. Y. Li, Sci. Rep. 5, 16108 (2015)}
\noindent{[33] J. L. Collins, A. Tadich, W. Wu, L. C. Gomes, J. N. B. Rodrigues, C. Liu, J. Hellerstedt, H. Ryu, S. Tang, S.-K. Mo, S. Adam, S. A. Yang, M. S. Fuhrer, and M. T. Edmonds, Nature 564, 390 (2018).}\\
\noindent{[34] P. E. Blochl, Phys. Rev. B 50, 17953 (1994).} \\
\noindent{[35] J. P. Perdew, K. Burke, and M. Ernzerhof, Phys. Rev. Lett. 77, 3865 (1996).} \\
\noindent{[36] S. L. Dudarev, G. A. Botton, S. Y. Savrasov, C. J. Humphreys, and A. P. Sutton, Phys. Rev. B 57, 1505 (1998).} \\
\noindent{[37] Q. S. Wu, S. N. Zhang, H. F. Song, M. Troyer, and A. A. Soluyanov, Comput Phys Commun 224, 405 (2018).} \\
\noindent{[38] M. P. L. Sancho, J. M. L. Sancho, and J. Rubio, J. Phys. F 15, 851 (1985).} \\
\noindent{[39] A. A. Mostofi, J. R. Yates, Y.-S. Lee, I. Souza, D. Vanderbilt,and N. Marzari, Comput. Phys. Commun. 178, 685 (2008).} \\
\noindent{[40] J. L. Lado, and J. Fernández-Rossier, 2D Materials 4, 035002 (2017).} \\
\noindent{[41] J. Liu, Q. Sun, Y. Kawazoe, and P. Jena. Phys. Chem. Chem. Phys. 18, 8777 (2016).} \\
\noindent{[42] W. B. Zhang, Q. Qu, P. Zhua, and C. H. Lam,  J. Mater. Chem. C 3, 12457 (2015).}\\
\noindent{[43] H. Zhang, W, Qin, M. Chen, P. Cui, Z. Zhang, and X. Xu,  Phys. Rev. B 99, 165410 (2019).}\\
\end{document}